\documentclass[runningheads]{llncs}
\usepackage{graphicx}
\usepackage{amsmath}
\usepackage{cite}
\begin{document}
\title{Constrained self-supervised method with temporal ensembling for fiber bundle detection on anatomic tracing data}
\titlerunning{Fiber bundle detection on macaque tracing data}
%
\author{Vaanathi Sundaresan\inst{1} \and 
Julia F. Lehman\inst{2} \and
Sean Fitzgibbon \inst{3} \and Saad Jbabdi\inst{3} \and 
Suzanne N. Haber\inst{2, 4} \and
Anastasia Yendiki\inst{1}}
\authorrunning{V. Sundaresan et al.}
%
\institute{Athinoula A. Martinos Center for Biomedical Imaging, Massachusetts General Hospital and Harvard Medical School, Charlestown, MA, United States \and
Department of Pharmacology and Physiology, University of Rochester School of Medicine, Rochester, NY, United States \and
Wellcome Centre for Integrative Neuroimaging, FMRIB Centre, Nuffield Department of Clinical Neurosciences, University of Oxford, UK \and
McLean Hospital, Belmont, MA, United States\\ \email{vsundaresan1@mgh.harvard.edu}}
\maketitle              
\begin{abstract}
Anatomic tracing data provides detailed information on brain circuitry essential for addressing some of the common errors in diffusion MRI tractography. However, automated detection of fiber bundles on tracing data is challenging due to sectioning distortions, presence of noise and artifacts and intensity/contrast variations. In this work, we propose a deep learning method with a self-supervised loss function that takes anatomy-based constraints into account for accurate segmentation of fiber bundles on the tracer sections from macaque brains. Also, given the limited availability of manual labels, we use a semi-supervised training technique for efficiently using unlabeled data to improve the performance, and location constraints for further reduction of false positives. Evaluation of our method on unseen sections from a different macaque yields promising results with a true positive rate of $\sim$0.90. The code for our method is available at \url{https://github.com/v-sundaresan/fiberbundle\_seg\_tracing} 

\keywords{Anatomic tracing \and fiber bundle detection \and self-supervised \and contrastive loss.}
\end{abstract}

\section{Introduction}
\label{sec:intro}
\begin{figure}[h!]
    \vspace{-2em}
    \centering
    \includegraphics[width=11cm]{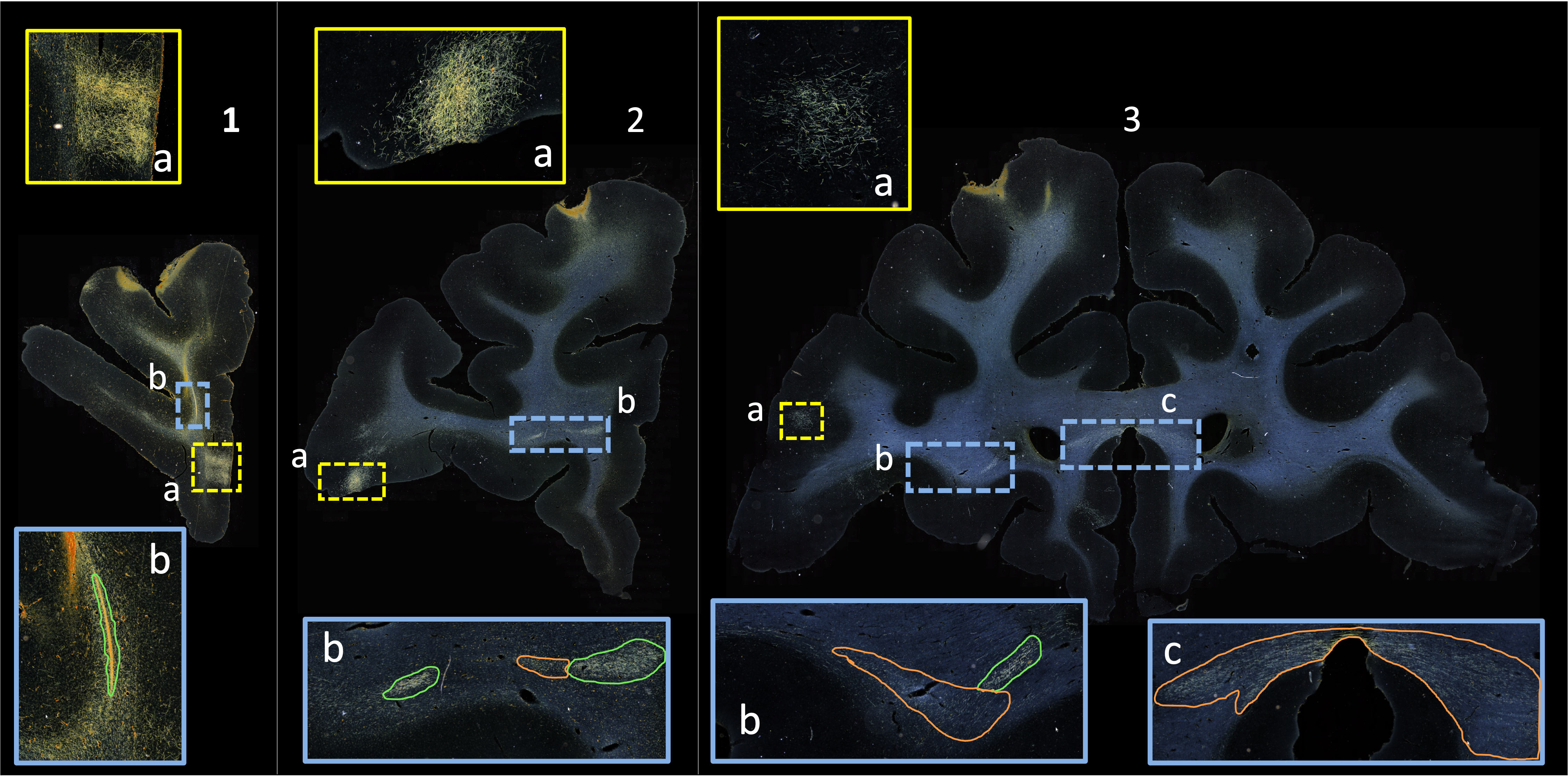}
    \vspace{-1em}
    \caption[]{\footnotesize{Photomicrographs showing coronal sections (1-3); (1a, 2a, 3a) terminal fields at different cortical locations. In the rostrocaudal direction, a fiber bundle stalk (1b) branches into two fiber bundles in prefrontal white matter (2b) and travels laterally in the external capsule (3b) and medially in the corpus callosum (3c). Manual chartings of dense and moderate bundles shown in green and orange respectively.} }
    \label{fig:intro}
    \vspace{-1em}
\end{figure}

Diffusion MRI (dMRI) allows us to probe the macroscopic organization and the microscopic features of white matter (WM) pathways \textit{in vivo}, and to study their role in psychiatric and neurological disorders \cite{yendiki2022post,grisot2021diffusion}. However, dMRI can only provide indirect measurements of axonal orientations based on water diffusion, and only at the mm scale. As a result, dMRI tractography sometimes fails, particularly in areas of complex fiber configurations, such as branching, fanning, or sharp turns \cite{grisot2021diffusion,maffei2022insights,schilling2019anatomical}. In contrast, anatomic tracing in non-human primates enables us to follow the trajectory of individual axons. As the fibers travel from an injection site, split into different fiber bundles and reach their terminal fields, they provide in-depth knowledge of how the brain is actually wired \cite{haber2022prefrontal,lehman2011rules,safadi2018functional,haynes2013organization}. Example tracer data from an injection site at the frontopolar cortex of a macaque monkey are shown in Fig.~\ref{fig:intro}. 
Anatomic tracing has been used to visualize tracts that are challenging for dMRI tractography \cite{maffei2022insights,safadi2018functional,jbabdi2013human}. For instance, WM fibers from prefrontal cortex travel through the gray matter of the striatum in small fascicles before entering the internal capsule (IC) \cite{safadi2018functional,jbabdi2013human}. From the IC, these fibers also enter anterior commissure perpendicular to the its main fiber tract orientation \cite{jbabdi2013human}. These discontinuities and tortuous trajectories of fibers confound dMRI tractography, but can be visualized clearly with anatomic tracing. 
However, 
the manual charting of fiber bundles on histological slides is extremely time consuming and labor-intensive, limiting the availability of annotated tracer data for large-scale validation studies.\\

The goal of this work is to accelerate this process by developing an accurate, automated method for detecting fiber bundles in histological sections. Existing work on automated fiber bundle segmentation on tracer data is quite scarce, and has been done primarily in marmoset brains \cite{woodward2020nanozoomer}. In contrast with these new world primates, the cortex of old world primates, (i.e. macaques) is evolutionarily much closer to humans and, as such the trajectory of fibers are more similar \cite{haber2022prefrontal,preuss2022evolution}. In general, automated fiber bundle detection on anatomic tracing data is challenging due to various confounding structures (e.g., terminal fields), sectioning distortions, localized and speckle artifacts and varying background intensities/textures. In this work, we propose a deep learning-based method for fully automated, accurate fiber bundle detection for the first time on tracer data from macaque brains, using only a few manually labeled sections. We use a multi-tasking model with anatomy-constrained self-supervised loss and utilise continuity priors to ensure accurate detection and to avoid false positives (FPs). So far, various semi/self-supervised and ensembling techniques have been shown to work well on noisy data and limited labels with uncertainties \cite{chen2020simple,wu2022cross,lai2021semi,perone2019unsupervised,huang2018omni}. Given the shortage of manually annotated tracer data, we successfully adapt a semi-supervised training technique to improve fiber bundle detection. We also evaluate the robustness of fiber bundle detection by validating our method on unseen sections from a different macaque brain.  

In addition to segmenting fiber bundles, our automated tool can also provide further quantification of fibers (e.g., density, volume). 
It is publicly available and we plan to deploy it for quantitative analyses of tracer data, and for large-scale validation studies where the accuracy of tractography algorithms will be evaluated across multiple seed areas.

\section{Method}
\label{sec:meth}
For training our method, we used (1) an encoder-decoder architecture for segmenting fiber bundles, while simultaneously discriminating fiber bundles from background using self-supervised contrastive loss, and (2) a temporal ensembling framework to efficiently use sections without manual charting from different brains.

\subsection{Self-supervised, temporal ensembling framework}
\label{ssec:selfsup_learn}
\begin{figure}[h!]
    \centering
    \vspace{-2em}
    \includegraphics[width=11cm]{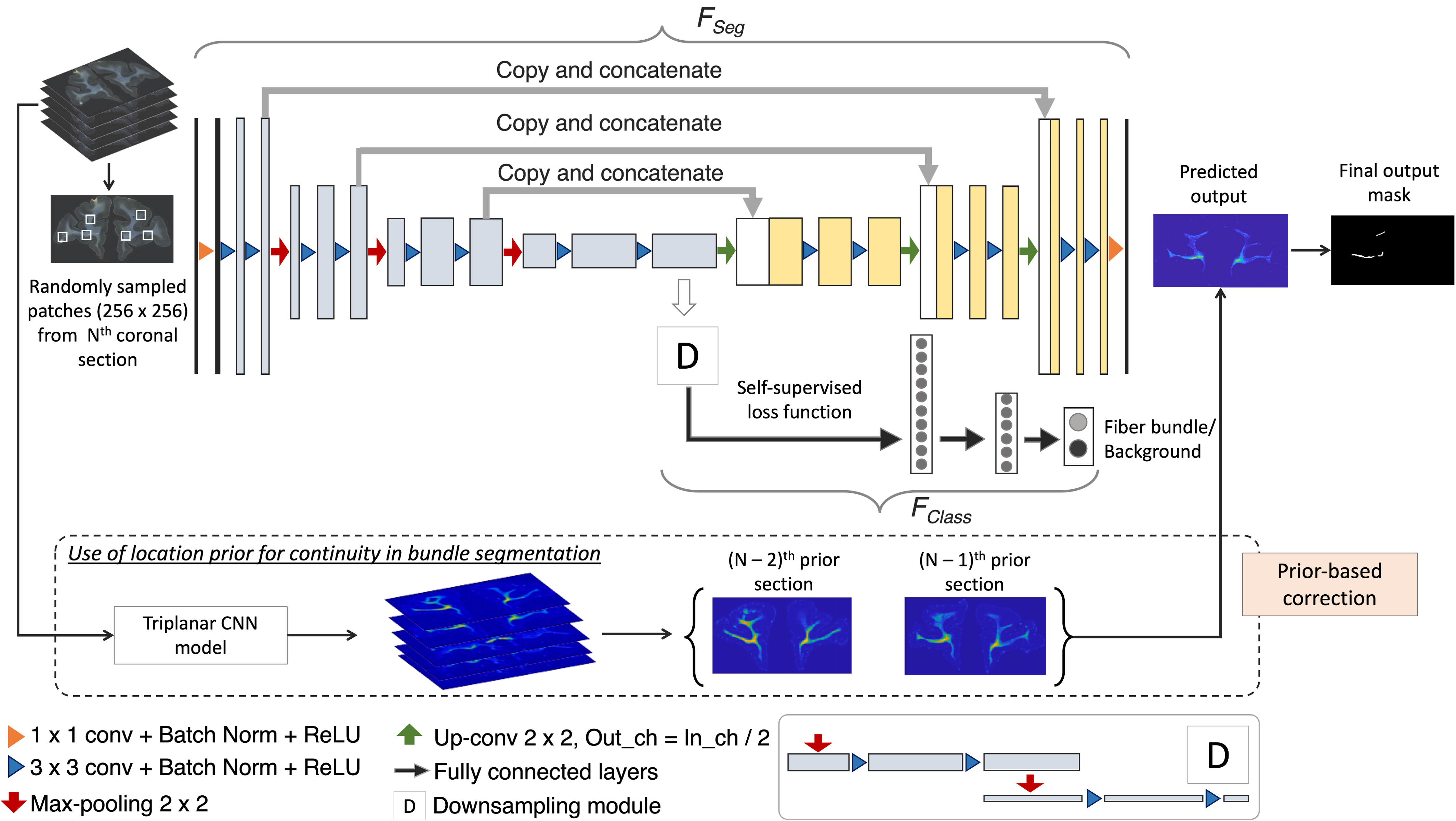}
    \vspace{-1em}
    \caption[]{\footnotesize{Network architecture used for fiber segmentation, and the use of location priors from rostral sections for false positive reduction. }}
    \label{fig:method}
    \vspace{-2em}
\end{figure}

\textbf{Anatomy-constrained self-supervised learning: }We used a 2D U-Net \cite{ronneberger2015u} to build a multi-tasking model as shown in Fig.~\ref{fig:method}, since U-Net is one of the most successful architectures for medical image segmentation tasks \cite{panayides2020ai}. The  multi-tasking model consists of a U-Net backbone ($F_{Seg}$) for segmenting the fiber regions/bundles and an auxiliary classification arm ($F_{Class}$) for discriminating fiber patches from background patches. We provided randomly sampled RGB patches of size 256 $\times$ 256 $\times$ 3 as input. 
$F_{Class}$ is connected to the bottleneck of the encoder of $F_{Seg}$, where the feature maps are passed through a downsampling module followed by two fully connected layers (\textit{fc}1024, \textit{fc}256) and an output layer with 2 nodes (fiber bundle vs background). The downsampling module consists of 2 max-pooling layers, each followed by two 3 $\times$ 3 convolution layers to extract high-level global features in the patches. We used focal loss (eqn.~\ref{eqn:focal}) for training $F_{Seg}$, since it handles class imbalance well \cite{lin2017focal}. The focal loss is given by:
\vspace{-1em}
\begin{equation}
\vspace{-1em}
FL(p_t) = -\alpha_t(1-p_t)^\gamma log(p_t), \quad p_t = 
\begin{cases}
    p, & \text{if} ~ y = 1 \\
    (1 - p), & \text{otherwise}
\end{cases} 
\label{eqn:focal}    
\end{equation}
where $\alpha$ and $\gamma$ are weighing and focusing parameters, respectively, and $p\in$[0,1] is the predicted probability for the fiber bundle class. As mentioned earlier, the manual charting of fiber bundles does not include all fiber regions and might not be precise along the boundaries. Moreover, we have texture variations and noise in the background. Therefore, we 
used a self-supervised technique for training $F_{Class}$ that learns intrinsic texture/intensity variations in addition to the fiber features from the manual charting alone. We used a contrastive loss function based on SimCLR \cite{chen2020simple}, where augmented data from each sample constitute the positive example to the sample while the rest were treated as negatives for the loss calculation. In SimCLR, random cropping and color distortions were shown to perform well. In our case we adapted the learning method by choosing augmentations better suited to our problem: (i) random cropping of patches closer to the input patch ($<$ 20$\mu$m), constrained within the white matter (by iterative sampling of patches until mean intensity criterion is satisfied), (ii) noise injection + Gaussian blurring (with randomly chosen $\sigma \in$ [0.05, 0.3]). The self-supervised loss with the above augmentations has two advantages: (1) effective separation between fiber and non-fiber background patches and (2) identification of fiber patches correctly even in the presence of artifacts, aided by the shared weights in the encoder of $F_{Seg}$. We used the contrastive loss \cite{chen2020simple} (eqn.~\ref{eqn:con_loss}) between positive pairs of patches $(i, j)$ of $F_{Class}$, given by:
\begin{equation}
\vspace{-0.5em}
CL(i, j) = \frac{exp(sim(f_i, f_j)/\tau)}{\sum_{k=1}^{2N} I_{k \neq i}~exp(sim(f_i, f_k)/\tau)}, \quad sim(x, y) = \frac{x^Ty}{|| x ||\hspace{0.25em} || y ||}
\label{eqn:con_loss}
\end{equation}
where $f$ is the output of $F_{Class}$, $sim(.)$ is the cosine similarity function, $I_{k \neq i} = 1$ if $k \neq i$, else 0 is the indicator function and $\tau$ is the temperature parameter. 

\hspace{-2em} \textbf{Temporal ensembling (TE) training: } Only $\sim$6\% of sections were manually charted, which would be insufficient for this challenging detection problem. Hence, after initially pretraining the model for $Np$ epochs using the manually charted samples alone, we used the additional unlabeled samples for training both $F_{Seg}$ and $F_{Class}$, using the temporal ensembling technique \cite{perone2019unsupervised}, where predictions from the previous $r$ epochs ($[P_{N-r}, ..., P_{N-1}]$) were averaged and thresholded to obtain the target label for the current epoch $N$ (we empirically set $r$ = 3). We used focal loss for pretraining the encoder-decoder of $F_{Seg}$, since contrastive loss was calculated at patch-level in $F_{Class}$. For the first 3 epochs after pretraining, predictions from the pretrained model $P_{N_p}$ were used for label generation. Averaging predictions reduced segmentation noise and aided in adapting the model to data from different brains. 

\textbf{Inference on test brain sections: } 
We obtained the predictions by applying the segmentation part $F_{Seg}$ of the model on the whole coronal sections (or patches of size 1024 $\times$ 1024 in the case of sections with dimensions larger than 1024 voxels).
\subsection{Continuity prior for false positive removal}
We used the spatial continuity of fibers from the injection site to remove obvious FPs (e.g., in cortical regions). We downsampled the sections by a factor of 10 and aligned the sections along the ventricles (or along the lateral edges of the brain for sections without ventricles) to roughly form 3D histological volumes. We applied a triplanar U-Net architecture used in \cite{sundaresan2021triplanar} to obtain a 3D priormap containing a crude segmentation of main dense fiber bundles and later upsampled it to the original dimensions. For each section, we computed the average of the segmented fiber bundle masks from the two nearest neighboring priormap sections (in rostral and/or caudal directions, if available). We removed any detected fiber bundle region in the current section, whose distance from the averaged bundles of prior sections was $>$0.2mm.\\
\textbf{Postprocessing: } We further reduced noisy regions and FPs by automated rejection of the predicted regions with area $<$2mm$^2$ and those near the brain outline.

\section{Experiments}
\label{sec:exps}

\textbf{Dataset used: } We used digitized, coronal histological sections from 12 macaques, with a slice thickness 50$\mu$m and in-plane resolution of 0.4$\mu$m. We considered every 8$^{th}$ section, resulting in a slice gap of 400$\mu$m (refer to \cite{lehman2011rules,haynes2013organization,haber2006reward} for more details on tracer injection, immunocytochemistry  and histological processing). Manual charting of fiber bundles labeled under `dense' and `moderate' bundles (examples shown in Fig.~\ref{fig:intro}) had been done previously by an expert neuroanatomist under dark-field illumination with a 4.0 or 6.4x objective, using Neurolucida software (MBF Bioscience). Manually charted region masks were registered with the tracing data using similarity (affine) transform with 6 DOF. 
Manual chartings were available for 2 macaques (61 sections). \textit{Dataset 1 (DS1)} consists of 465 sections, including 25 charted sections (out of 61) from one macaque and 440 unlabeled sections from 10 macaques for training. \textit{Dataset 2 (DS2)} consists of 36 charted sections from the other annotated macaque for testing. Both datasets were downsampled in-plane by a factor of 4 for training and testing.\\

\textbf{Implementation details: }For training, we used the Adam optimizer \cite{kingma2014adam} ($\epsilon = 10^{-3}$), batch size = 8, pretraining epochs ($N_p$) = 100 and trained with TE for 100 epochs with a patience value of 25 epochs for early stopping (converged at $\sim$90 epochs). For focal loss, we used $\alpha =$ 0.25; $\gamma =$ 2, and for contrastive loss, we used $\tau =$ 0.5. The hyperparameters were chosen empirically. For $F_{Seg}$, we augmented data using translation (offset $\in$ [-50, 50] voxels), rotation ($\theta \in$ [-20$^o$, 20$^o$]), horizontal/vertical flipping and scaling ($s \in$ [0.9, 1.2]). The model was implemented using PyTorch 1.10.0 on Nvidia GeForce RTX 3090, taking $\sim$10 mins/epoch for $\sim$22,000 samples (training:validation = 90:10).\\

\textbf{Experimental setup and evaluation metrics: } We performed 5-fold cross-validation on 465 sections (440 unlabeled + 25 labeled) from DS1 with a training-validation-testing split ratio of 80-13-5 sections (for each fold, only manually charted sections were used for testing). We then trained the model on the DS1 and tested it on the unseen dataset DS2 (sections from a macaque different from the training one). We also performed an ablation study of the method on DS2. We studied the detection performance for dense and moderate bundles with the addition of individual components of the method: (i) $F_{Seg}$ with cross-entropy loss (CE loss), (ii) $F_{Seg}$ with focal loss, (iii) $F_{Seg}$ with addition of $F_{Class}$ with contrastive loss (focal loss + ss\_con loss), (iv) $F_{Seg}$ and $F_{Class}$ with TE (focal loss + ss\_con loss + TE). We used the same postprocessing for all cases (i-iv), since our main aim is to study the effect of addition of $F_{Seg}$, ss\_con loss and TE, rather than postprocessing. 
For evaluation, we used the following metrics: (1) True positive rate (TPR): number of true positive bundles / total number of true bundles charted manually, (2) Average number of FPs ($FP_{avg}$): number of false positive bundles / number of test sections and (3) Fiber density ratio ($fib\_dens$): ratio between fiber voxels (obtained from fiber binary map) and the total bundle area. We obtained the fiber binary map by considering fibers within the bounding box of the bundle, enhancing the contrast using contrast-limited adaptive histogram equalization \cite{zuiderveld1994contrast} and thresholding at the 95$^{\text{th}}$ percentile of intensity values (sample fiber maps shown in Fig.~\ref{fig:visres}). We calculated the difference in the ratio ($\delta_{fib\_dens}$) between the manual charting and the detected bundles. 

\section{Results and discussion}
\label{sec:res_disc}
\begin{figure}[h!]
    \centering
    \includegraphics[width=11cm]{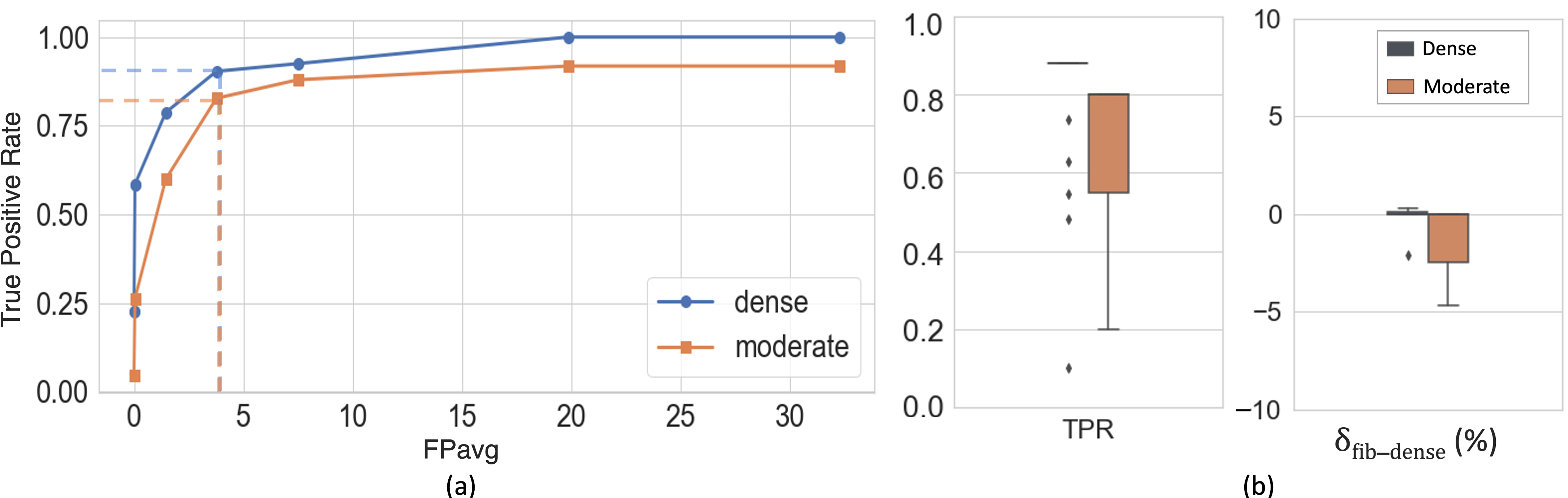}
    \vspace{-0.5em}
    \caption[]{\footnotesize{Results of cross-validation on DS1. (a) FROC curves for fiber bundle detection. (b) boxplots of TPR and $\delta_{fib\_dens}$ at $FP_{avg}$=2 FPs/section, after applying continuity constraints and postprocessing.}}
    \label{fig:roc_box}
\end{figure}

\textbf{Cross-validation (CV) on DS1: }On performing 5-fold CV on DS1, we obtained better performance in the detection of dense bundles than moderate ones due to the contrast, increased fiber density and texture differences of the former with respect to the background. Fig.~\ref{fig:roc_box}(a) shows FROC curves for dense and moderate fiber bundle detection. We obtained a TPR of 0.92/0.84 for dense/moderate bundles at 3.7 FPs/section at the \textit{elbow point} (shown in dotted lines) for a threshold value of 0.4. 

\begin{table*}[h!]
\centering
\scriptsize
\caption{Results of 5-fold cross-validation and ablation study. CE loss - Cross-entropy loss, Focal loss - $F_{Seg}$ with focal loss, ss\_con loss - $F_{Class}$ with self-supervised contrastive loss, TE - temporal ensembling. (*) indicates significant improvements in the results compared to the previous row, determined using paired two-tailed T-tests. The best performance in the ablation study is highlighted in bold. $\uparrow$/$\downarrow$ indicate that higher/lower values lead to better results.}
{\renewcommand{\arraystretch}{1.5}%
\begin{tabular}{l|cc|cc|c}
  \hline
  \parbox[c][][t]{3.0cm}{\raggedright }& 
  \multicolumn{2}{|c|}{\parbox[c][][t]{3.4cm}{\centering \textbf{TPR $\uparrow$}}}& 
  \multicolumn{2}{|c|}{\parbox[c][][t]{3.4cm}{\centering \textbf{$|\mathbf{\delta_{fib\_dens}}|$ (\%) $\downarrow$}}} & 
  \multicolumn{1}{|c}{\parbox[c][][t]{1.7cm}{\centering \textbf{$\mathbf{FP_{avg}}$ $\downarrow$}}}  \\
  \cline{2-5}
  \parbox[c][][t]{3.0cm}{\raggedright }& 
  \parbox[c][][t]{1.7cm}{\centering \textbf{Dense}} &
  \parbox[c][][t]{1.7cm}{\centering \textbf{Moderate}} &
  \parbox[c][][t]{1.7cm}{\centering \textbf{Dense}} &
  \parbox[c][][t]{1.7cm}{\centering \textbf{Moderate}} &
  \parbox[c][][t]{1.7cm}{\centering \textbf{}} \\
  \hline
  \parbox[c][][t]{3.0cm}{\raggedright 5-fold cross-validation}& 
  \parbox[c][][t]{1.7cm}{\centering 0.90$\pm$0.20} &
  \parbox[c][][t]{1.7cm}{\centering 0.82$\pm$0.17} &
  \parbox[c][][t]{1.7cm}{\centering 1.5$\pm$0.45} &
  \parbox[c][][t]{1.7cm}{\centering 3.5$\pm$1.67} &
  \parbox[c][][t]{1.7cm}{\centering 2.0} \\
  \hline
  \multicolumn{6}{c}{\parbox[c][][t]{10.9cm}{\centering \textbf{Ablation study}}}\\
  \hline
  \parbox[c][][t]{3.0cm}{\raggedright CE loss}& 
  \parbox[c][][t]{1.7cm}{\centering 0.76$\pm$0.33} &
  \parbox[c][][t]{1.7cm}{\centering 0.65$\pm$0.29} &
  \parbox[c][][t]{1.7cm}{\centering 4.1$\pm$0.39} &
  \parbox[c][][t]{1.7cm}{\centering 5.8$\pm$1.09} &
  \parbox[c][][t]{1.7cm}{\centering 7.5} \\
  \parbox[c][][t]{3.0cm}{\raggedright Focal loss}& 
  \parbox[c][][t]{1.7cm}{\centering *0.85$\pm$0.31} &
  \parbox[c][][t]{1.7cm}{\centering *0.71$\pm$0.34} &
  \parbox[c][][t]{1.7cm}{\centering *3.0$\pm$0.48} &
  \parbox[c][][t]{1.7cm}{\centering *4.6$\pm$1.01} &
  \parbox[c][][t]{1.7cm}{\centering *4.0} \\
  \parbox[c][][t]{3.0cm}{\raggedright Focal loss + ss\_con loss}& 
  \parbox[c][][t]{1.7cm}{\centering *0.88$\pm$0.20} &
  \parbox[c][][t]{1.7cm}{\centering *0.78$\pm$0.23} &
  \parbox[c][][t]{1.7cm}{\centering 2.8$\pm$0.21} &
  \parbox[c][][t]{1.7cm}{\centering *4.0$\pm$0.86} &
  \parbox[c][][t]{1.7cm}{\centering 3.5} \\
  \parbox[c][][t]{3.0cm}{\raggedright Focal loss + ss\_con loss + TE \vspace{2pt}}& 
  \parbox[c][][t]{1.7cm}{\centering \textbf{*0.89$\pm$0.21}} &
  \parbox[c][][t]{1.7cm}{\centering \textbf{0.79$\pm$0.30}} &
  \parbox[c][][t]{1.7cm}{\centering \textbf{*2.0$\pm$0.21}} &
  \parbox[c][][t]{1.7cm}{\centering \textbf{*3.7$\pm$0.80}} &
  \parbox[c][][t]{1.7cm}{\centering \textbf{*2.5}} \\
  \hline
  \end{tabular}}
\label{table:abl_study}
\end{table*}
\begin{figure}[h!]
    \centering
    \includegraphics[width=\textwidth]{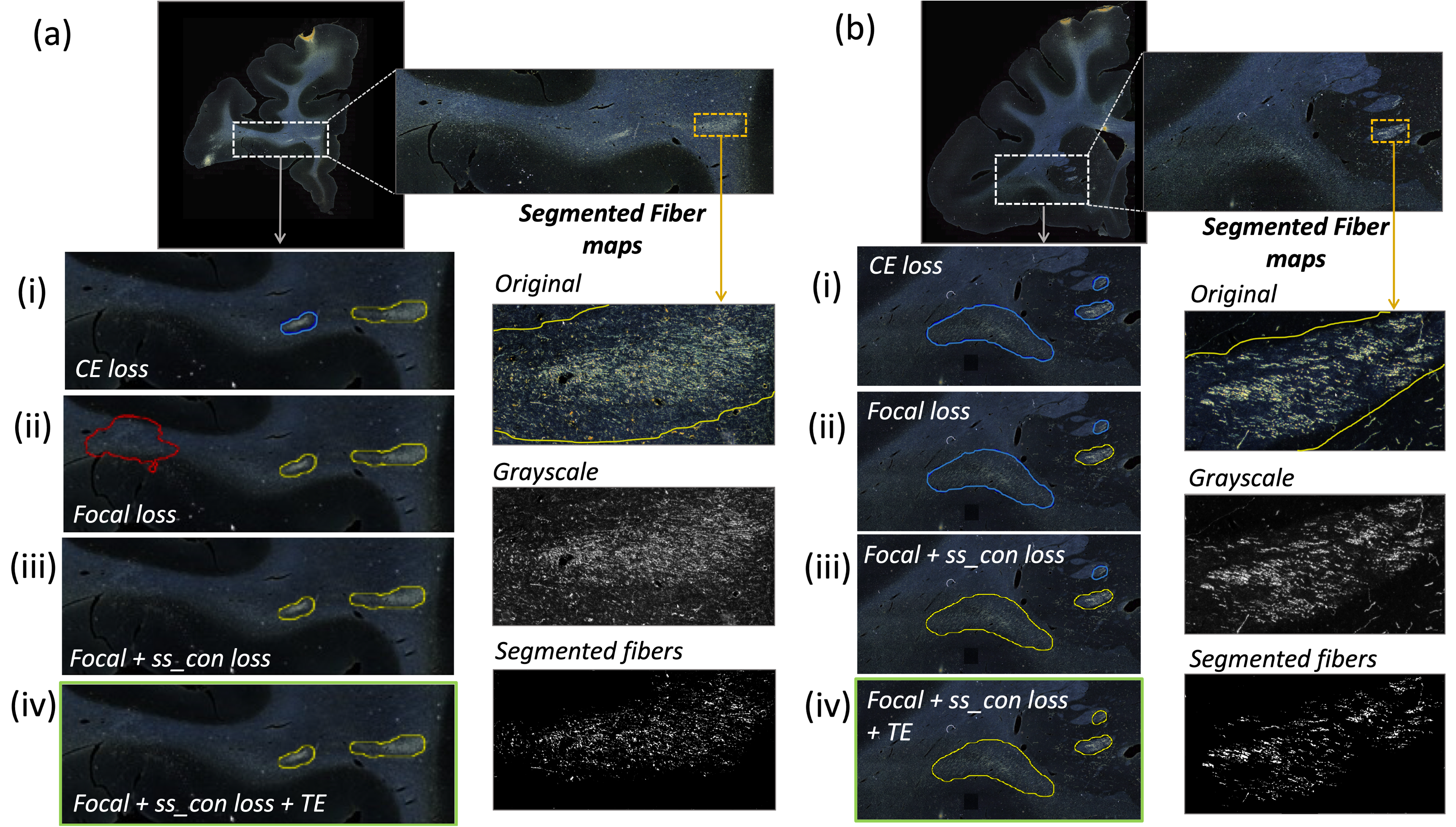}
    \vspace{-2em}
    \caption[]{\footnotesize{Two sample results of the ablation study, with the profile of fibers within detected bundles. (a, b) Sections with ROIs enlarged (white dotted box); (i – iv) Ablation study results on the ROIs - TP, FP and FN bundles shown in yellow, red and blue outlines respectively (the proposed method highlighted in green box (iv)). Further enlarged ROIs (orange dotted box) containing fibers in the original RGB, grayscale and fiber binary maps.}}
    \label{fig:visres}
    \vspace{-2em}
\end{figure}
Fig.~\ref{fig:roc_box}(b) shows the boxplots of TPR and $\delta_{fib\_dens}$ values after postprocessing (performance values reported in Table~\ref{table:abl_study}). We observed a significant reduction of $FP_{avg}$ ($p$ $<$ 0.05) after postprocessing, mainly due to continuity constraints, for much lower changes in TPR values.  Typically, fiber density ratios $fib\_dens$ ranged between $\sim$6-20\% and $\sim$2-10\% for dense and moderate bundles respectively. We obtained mean $\delta_{fib\_dens}$ = -1.5\% and -3.5\% for dense and moderate bundles, respectively. While \% values closer to 0 are better, the negative values indicate more fibers in the predicted regions than manual charting in most cases, indicating that predicted regions recover most of the fibers within the bundle. 

\textbf{Ablation study results on DS2: } We trained the method on dataset DS1 for ablation study cases (i-iv), tested on DS2 (from different brain) and used a threshold of 0.4 to obtain binary maps. We used the same postprocessing for all cases (i-iv) of the study. Table~\ref{table:abl_study} reports the results of the study and Fig.~\ref{fig:visres} shows sample results of the ablation study for a dense bundle in the prefrontal white matter (a) and a moderate bundle in the IC (b). Among all the methods, experiments using focal loss (ii - iv) gave significantly better performance than the CE loss (case i), showing that focal loss was better at handling the heavy class imbalance. Also, using the self-supervised contrastive loss (ss\_con loss) significantly improved TP regions (especially moderate fiber bundles) and reduced $FP_{avg}$ due to the better discrimination between subtle variations in the background intensity and texture. We also observed a significant reduction in $\delta_{fib\_dens}$ for focal loss + ss\_con loss (case iii) in moderate fiber bundles (where the fibers are sparser than dense bundles). This shows that contrastive loss function not only reduced FPs, but also improved the segmentation of predicted regions. Using TE (case iv) further improved the detection, especially increasing the TPR of dense bundles and reducing $FP_{avg}$. We observed that the value of $r$ (number of prior epochs to predict the target labels) in TE played a crucial role in the reduction of prediction noise. We set $r$ = 3 because it significantly reduced $FP_{avg}$ over $r$ = 1 ($p <$ 0.01), but provided $FP_{avg}$ values not significantly different from those with  higher $r$ = 5 ($p =$ 0.52).


The main source of FPs included terminal fields (shown in fig~\ref{fig:intro}), artifacts such as glare or dust particles, and other structures with similar intensity profiles. Use of continuity priors and ss\_con loss was highly useful in removing these spurious regions. Currently, inclusion of such priors in the training framework was not possible due to the lack of sufficient number of manual chartings for consecutive sections. 
Hence, a future direction of this work could explore the possibility of integrating the priors within the training framework for further reduction of FPs, and improving the method to reduce the variation (indicated by standard deviation) in our results. 
Another area for further study is the quantification of fiber-level characteristics (e.g., fiber density and orientation).

\section{Conclusions}
\label{sec:conc}
In this work, we proposed an end-to-end automated, anatomy-constrained self supervised learning tool for accurate detection of fiber bundles on macaque tracer data. With only $\sim$6\% of training data manually charted, we achieved TPR of 0.90/0.80 for dense/moderate fiber bundles on different macaque brain sections. Our tool could be used for generating voxel visitation maps to analyse the precise route of axon bundles and their densities along fiber trajectories for voxel-level validation of dMRI tractography across multiple seed regions. The code for our method is available at \url{https://github.com/v-sundaresan/fiberbundle\_seg\_tracing}. 
    
\section*{Acknowledgements}
\label{sec:ack}
This work was supported by the National Institute of Mental Health (R01-MH045573, P50-MH106435), the National Institute of Neurological Disorders and Stroke (R01-NS119911), and the National Institute of Biomedical Imaging and Bioengineering (R01-EB021265).
%
%
%
%

\end{document}